\documentclass[eqsecnum,showpacs,showkeys,floats,aps,nofootinbib,preprint]{revtex4}
\usepackage{bm,amsfonts, mathtools}
\usepackage{graphicx,color, wasysym}
\usepackage{textcomp}
\usepackage{amsmath,amssymb,latexsym,epsfig}
\def\nn{\nonumber}

\begin{document}

\makeatletter

\title{Relativistic harmonic oscillator}

\author{D. Babusci}
\email{danilo.babusci@lnf.infn.it}
\affiliation{INFN - Laboratori Nazionali di Frascati, via E. Fermi, 40, IT 00044 Frascati (Roma), Italy}

\author{G. Dattoli}
\email{dattoli@frascati.enea.it}
\author{M. Quattromini}
\email{quattromini@frascati.enea.it}
\author{E. Sabia}
\email{sabia@frascati.enea.it}
\affiliation{ENEA - Centro Ricerche Frascati, via E. Fermi, 45, IT 00044 Frascati (Roma), Italy}

\begin{abstract}
We consider the relativistic generalization of the harmonic oscillator problem by addressing different questions 
regarding its classical aspects. We treat the problem using the formalism of Hamiltonian mechanics. A Lie 
algebraic technique is used to solve the associated Liouville equations, yielding the phase space evolution of an 
ensemble of relativistic particles, subject to a ``harmonic" potential. The non-harmonic distortion of the spatial 
and momentum distributions due to the intrinsic non-linear nature of the relativistic contributions are discussed. 
We analyze the relativistic dynamics induced by two types of Hamiltonian, which can ascribed to those of harmonic 
oscillators type. Finally, we briefly discuss the quantum aspects of the problem by considering possible strategies 
for the solution of the associated Salpeter equation.
\end{abstract}

\maketitle

\section{Introduction}\label{s:intro}
The relativistic harmonic oscillator (RHO) is a topic of fundamental importance in Physics. Although widely discussed in the past (see, for example Ref. \cite{Moreau}) 
it still contains elements of interest, deserving careful comments, both in classical and quantum mechanics. The spinless Salpeter equation with a quadratic potential 
cannot be solved in analytical form \cite{Lucha} and the corrections induced by the relativistic effect in the phase space evolution of an ensemble of particles ruled by 
a harmonic potential have not been discussed yet. This point is raised by a genuine practical necessity, emerging in charged beam transport in magnetic systems. 
In particular, in the case of propagation through quadrupole lenses, the beam is ruled by a potential of harmonic oscillator type \cite{Drago}. It may therefore 
happen that, in absence of a clear distinction, phase space distortions due to relativistic corrections could be misinterpreted and ascribed, for example, to quadrupole 
aberrations. The dynamics of the relativistic harmonic oscillator is, indeed, not harmonic at all. The relativistic nature of the problem induces non-linear corrections 
which determines a genuine anharmonic behavior.

In this paper we will consider the classical aspects of the RHO and we will treat the problem \textit{ab initio}. Just to give an idea of the problems we are going to treat, 
we consider the case of a relativistic particle subject to a linear potential whose Hamiltonian is 
\begin{equation}
\label{Ham}
H = c\,\sqrt{p^2 + (m_0\,c)^2} - F\,x, 
\end{equation}
and the associated equations of motion 
\begin{equation}
\dot{x} = \frac{\partial H}{\partial p} =  \frac{p\,c}{\sqrt{p^2 + (m_0\,c)^2}}, \qquad\qquad 
\dot{p} = - \frac{\partial H}{\partial x} = F = m_0\,a,
\end{equation}
are readily integrated as
\begin{equation}
x (t) = x_0 + \left(\sqrt{1 + \Pi^2} - \sqrt{1 + \Pi_0^2}\right)\,\frac{c^2}{a}, \qquad\qquad 
\Pi (t) =  \frac{p (t)}{m_0\,c} = \Pi_0 + \frac{a\,t}{c}.
\end{equation}
The non-relativistic limit of these equations is given by
\begin{equation}
x^{\mathrm{(NR)}}  (t) = x_0 + \frac12\,\left[\left(\Pi_0 + \frac{a\,t}{c}\right)^2 - \Pi_0^2\right]\,\frac{c^2}{a}, \qquad\qquad 
\Pi^{\mathrm{(NR)}} (t) = \Pi_0 = \frac{p_0}{m_0\,c} = \gamma_0\,\beta_0,
\end{equation}
($\beta = v/c, \gamma = 1/\sqrt{1 - \beta^2}$) and is recognized as the classical equation for the uniformly accelerated motion.

It has been noted by Harvey \cite{Harvey} that the derivation, in an unambiguously Lorentz invariant fashion, of the Hamiltonian of a relativistic particle subject to a scalar 
potential, hinges on the relaxation of the requirement that the rest mass be a constant. The procedure proposed in Ref. \cite{Harvey} is reminiscent of the Lorentz covariant 
theory of gravitation (see last reference in \cite{Harvey}), where the rest mass is potential dependent. By adhering to this point of view, an alternative relativistic Hamiltonian 
with a linear potential can be written, according to the prescription
\begin{equation}
\label{Massred}
H^2 - (p\,c)^2 = (m\,c^2)^2,\qquad\qquad m = m_0 - \frac{F\,x}{c^2}
\end{equation}
where it is evident that the potential redefines the particle mass, and the following Hamiltonian is obtained
\begin{equation}
\label{Ham2}
H = \sqrt{(p\,c)^2 + (m_0\,c^2 - F\,x)^2}.
\end{equation}
The physics described by this Hamiltonian totally differs from that of Hamiltonian \eqref{Ham}. For example, the latter can be exploited to study the equation of motion 
of a relativistic charged particle subject to an external electric field, while Eq. \eqref{Ham2} could be viewed as that relevant to a relativistic charged particle in a constant 
magnetic field\footnote{The relativistic Hamiltonian for a charged particle with charge $q$ and mass $m_0$ moving in a magnetic field with intensity $B$ directed along the 
$z$-direction can be written as $H = c\,\sqrt{p_x^2 + (p_y - q\,B\,x)^2 + (m_0\,c)^2}$, mathematically equivalent to Eq. \eqref{Ham2}.}. 

To better appreciate the differences, we make a comparison between the solutions of the respective equations of motion, which in the case of the Hamiltonian \eqref{Ham2}, 
read
\begin{equation}
\label{eqmot}
\dot{x} = \frac{p\,c^2}{H}, \qquad\qquad \dot{p} = \frac{m_0\,c^2 - F\,x}{H}\,F
\end{equation}
and can be integrated to get 
\begin{equation}
x (t ) = \frac{c^2}{a}\,\left\{1 + \gamma_0\,\left(\sqrt{1 - \delta^2}\,\sin \xi - \delta\,\cos \xi\right)\right\} ,
\end{equation} 
with
\begin{equation}
\xi = \frac{a\,t}{\gamma_0},\qquad\qquad 
\delta = \frac1{\gamma_0}\,\left(1 - \frac{a}{c^2}\,x_0\right). \nn
\end{equation}
The motion, as also evident from Eq. \eqref{eqmot}, is essentially harmonic with a characteristic frequency $\omega = a/(\gamma_0\,c)$. It is interesting to note that, the solution 
(including the velocity) can be written in the form of a rotation, namely 
\begin{equation}
\label{Xcap}
\left(\begin{array}{c}
X (\xi) \\
X^\prime (\xi)
\end{array}
\right) = \left(
\begin{array}{cc}
\sin \xi & - \cos \xi \\
\cos \xi & \sin \xi
\end{array}\right)\,\left(
\begin{array}{c}
X_0 \\
X_0^\prime
\end{array}
\right),
\end{equation}
where
\begin{equation}
X (\xi) = \frac1{\gamma_0}\,\left[x (\xi) - \frac{c^2}{a}\right],
\end{equation}
and, thus, $X_0 = \sqrt{1 - \delta^2}$, $X_0^\prime = \delta$. From this result we see that the particle motion is essentially oscillatory, and for short times 
($t \ll c\,\gamma_0/a$) the non-relativistic equations of the uniformly accelerated motion is obtained.

In the forthcoming sections we will discuss analogous problems in the case of the harmonic potential. We will address our analysis by studying the Liouville equation 
describing the evolution of phase-space distributions of an ensamble of relativistic particles.

\section{The Liouville equation for the relativistic harmonic oscillator}
Let us now consider the following Hamiltonian
\begin{equation}
H = c\,\sqrt{p^2 + (m_0\,c)^2} + V (x)
\end{equation}
which rules the relativistic evolution of a particle with mass $m_0$ subject to a potential $V (x)$. The solution of this dynamical problem is accomplished by solving 
the associated Hamiltonian equations, that introducing the notation $Z = (x, p)^T$, can be written as
\begin{equation}
\label{Zeq}
\dot{Z} = H \circ Z,
\end{equation}
where
\begin{equation}
H \circ F (x, p) = \{H, F\} = \frac{\partial H}{\partial p} \frac{\partial F}{\partial x} - \frac{\partial H}{\partial x} \frac{\partial F}{\partial p}.
\end{equation}

The solution of Eq. \eqref{Zeq} can be formally expressed as the following Lie series
\begin{equation}
Z = U_H (t)\,Z_0 = \sum_{n = 0}^\infty \frac{t^n}{n!}\,(H \circ)^n\,Z_0 
\end{equation}
where $U_H (t) = e^{t H \circ}$ is the Hamilton evolution operator, and the following notation has been used
\begin{equation}
(H \circ)^0\,F = F, \qquad (H \circ)\,F = \{H, F\}, \qquad (H \circ)^2\,F = \{H, \{H, F\}\}, \qquad \dots
\end{equation}
This procedure is reminiscent of analogous methods adopted in quantum mechanics to treat the time-dependent Schr\"odinger equation \cite{Dattoli97}. The problem 
is not in general amenable for analytical solutions. For example, in classical (non-relativistic) mechanics exact solutions can be obtained only in case of quadratic Hamiltonians. 
In the relativistic case even quadratic potentials are not amenable for an exact solution. In these situations, the search of solutions of approximate nature is the only way 
to address the problem. The symmetric exponential split rule has been proven an efficient tool. According to it, we can write
\begin{equation}
e^{t H \circ} = e^{(t/2) H_1 \circ}\,e^{t H_2 \circ}\,e^{(t/2) H_1 \circ} + O (t^3)
\end{equation}
where 
\begin{equation}
H_1 = c\,\sqrt{p^2 + (m_0\,c)^2}, \qquad\qquad H_2 = V (x).
\end{equation}
The solution of the problem is then obtained by a repeated application of the evolution operator calculated for short time interval steps of duration $\delta t = t/N$, namely
\begin{equation}
\label{Ziter}
Z_{n + 1} = e^{\delta t\,H \circ}\,Z_n \qquad\qquad (n = 0, 1, \dots, N - 1).
\end{equation} 

Since we are interested in particle ensemble distributions ruled by relativistic Hamiltonians, we can adopt a complementary procedure based on the solution of the Liouville equation 
associated with the Hamiltonian $H$. It describes the evolution of the particles phase space density distribution $\rho (x, p; t)$ and reads ($t$ is time in the laboratory frame)
\begin{equation}
\label{Liou}
\frac{\partial}{\partial t} \rho (x, p; t) = \hat{L}\,\rho (x, p; t)
\end{equation}
whose solution can formally be written as
\begin{equation}
\rho (x, \Pi; t) = \hat{U}_L (t)\,\rho (x, \Pi; 0) \qquad\qquad \hat{U}_L (t) = e^{\,t\,\hat{L}}
\end{equation}
where the Liouville operator $\hat{L}$ is given by
\begin{equation}
\hat{L} = \left[\frac{\partial H}{\partial x}\frac{\partial}{\partial p} - \frac{\partial H}{\partial p}\frac{\partial}{\partial x}\right].
\end{equation}
The use of well-known techniques from operator calculus \cite{Dattoli97} will be exploited in the forthcoming section to discuss the problems concerning the relativistic phase space 
conservation. Here we just remind that the function $\rho (x, p; t)$ transforms as a Lorentz scalar and its integral over the phase space variable is normalized to unity.

In the case of free relativistic particles the Liouville operator is\footnote{We have rescaled the phase space coordinates and exploited the adimensional momentum $\Pi$ introduced 
in the previous section.}
\begin{equation}
\hat{L} = - c\,\frac{\Pi}{\sqrt{1 + \Pi^2}}\,\frac{\partial}{\partial x}
\end{equation}
and, thus, the action of the associated evolution operator on the initial distribution reduces to a shift of the spatial coordinate, namely
\begin{align}
\rho (x, \Pi; t) &= \exp\left\{- c\,t\,\frac{\Pi}{\sqrt{1 + \Pi^2}}\,\frac{\partial}{\partial x}\right\}\,\rho (x, \Pi; 0) \nn \\
& = \rho\left(x - c\,t\,\frac{\Pi}{\sqrt{1+\Pi^2}}, \Pi; 0\right).
\end{align}
The Liouville operator associated with the Hamiltonian \eqref{Ham} is
\begin{equation}
\hat{L} = - c\,\frac{\Pi}{\sqrt{1 + \Pi^2}}\,\frac{\partial}{\partial x} - \frac{a}{c}\,\frac{\partial}{\partial\Pi}. 
\end{equation}
and the evolution of the corresponding phase space distribution function writes 
\begin{equation}
\label{2expl}
\rho (x, \Pi; t) = \rho \left(x + \frac{c^2}{a}\,\left[\sqrt{1 + \left(\Pi - \frac{a\,t}{c}\right)^2} - \sqrt{1 + \Pi^2}\right], \Pi - \frac{a\,t}{c}; 0\right).
\end{equation}
We will use this equation as a benchmark of an approximate solution based on the split operator method, then applied to the case of more complicated potentials for which no exact 
solution is available. We start decomposing the Liouville operator in two parts
\begin{equation}
\hat{L}_1 =  \frac{a}{c}\,\frac{\partial}{\partial\Pi}, \qquad\qquad 
\hat{L}_2 = - c\,\frac{\Pi}{\sqrt{1 + \Pi^2}}\,\frac{\partial}{\partial x}. 
\end{equation}
Since $\hat{L}_1$ and $\hat{L}_2$ do not commute, the evolution operator cannot be naively disentangled, the use of the same procedure leading to Eq. \eqref{Ziter}, yields 
\begin{equation}
\rho_{n + 1} = \hat{U}_L\,\rho_{n} \qquad\qquad (n = 0, 1, \dots, N - 1) 
\end{equation}
with
\begin{equation}
\hat{U}_L \cong \exp\left(\frac{\delta t}2\,\hat{L}_1\right)\,\exp(\delta t\,\hat{L}_2)\,\exp\left(\frac{\delta t}2\,\hat{L}_1\right).
\end{equation}
By repeatedly acting with this operator on the initial function $\rho_0 = \rho(x_0, \Pi_0; 0)$, and taking into account the shift of the coordinates $x$ and $\Pi$ induced at each step by 
the exponential operator, we end up with
\begin{equation}
\rho_n = \rho(x_n, \Pi_n; 0)
\end{equation}
where
\begin{equation}
x_n = x_{n - 1} - \frac{\Pi_n - \displaystyle \frac{a}{2\,c}\,\delta t}
{\sqrt{1 + \left(\Pi_n - \displaystyle \frac{a}{2\,c}\,\delta t\right)^2}}\,c\,\delta t, \qquad\qquad
\Pi_n = \Pi_{n - 1} - \frac{a}{c}\,\delta t.
\end{equation}
An advantage of the symmetric split is that of providing at any step a norm preserving distribution. The comparison between the exact solution and the one obtained with this 
technique has been completely satisfactory, and therefore the method can be safely applied to the case of a quadratic potential. 

\begin{figure}[htb]
\includegraphics[scale=0.42]{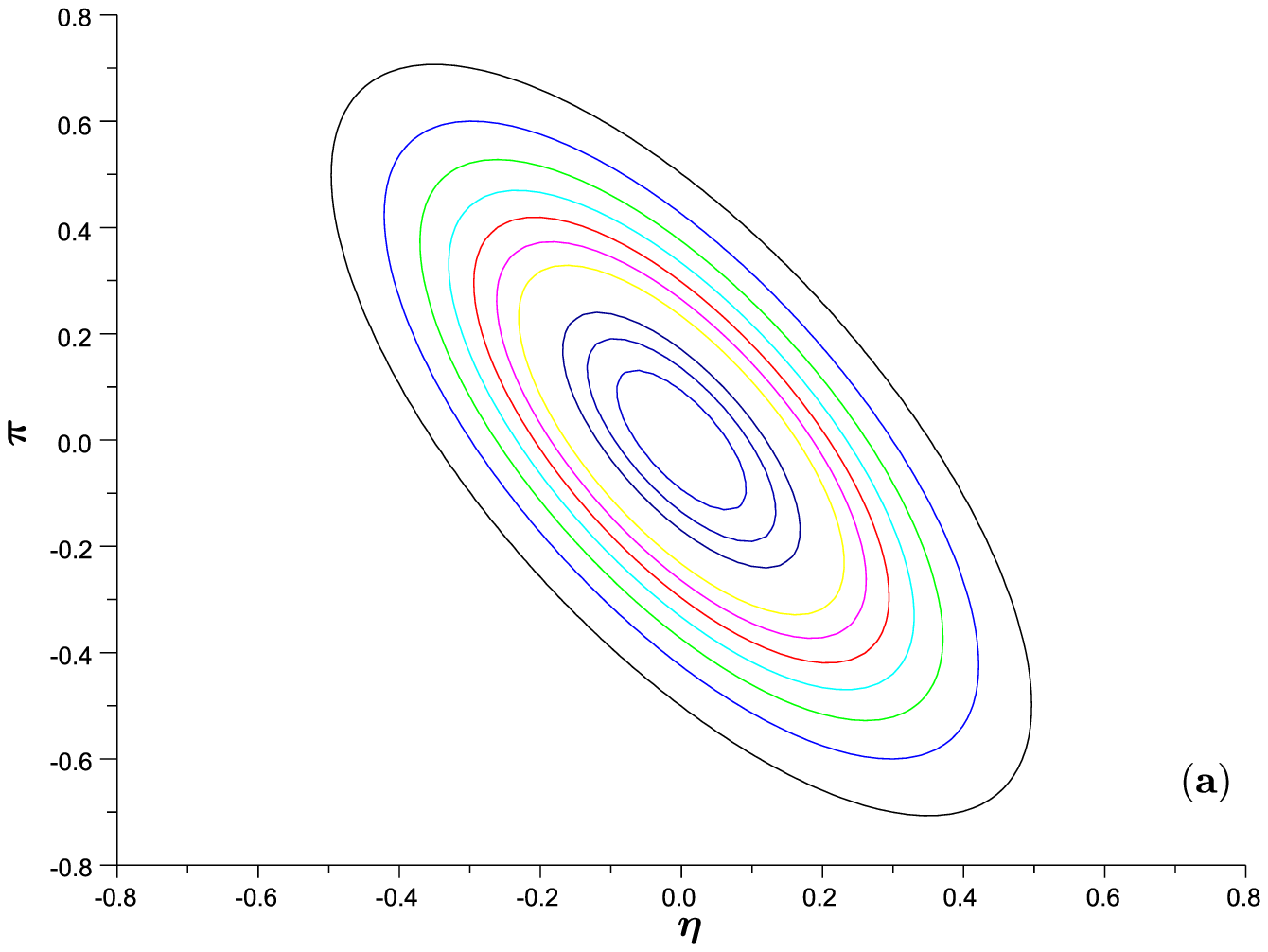}\includegraphics[scale=0.42]{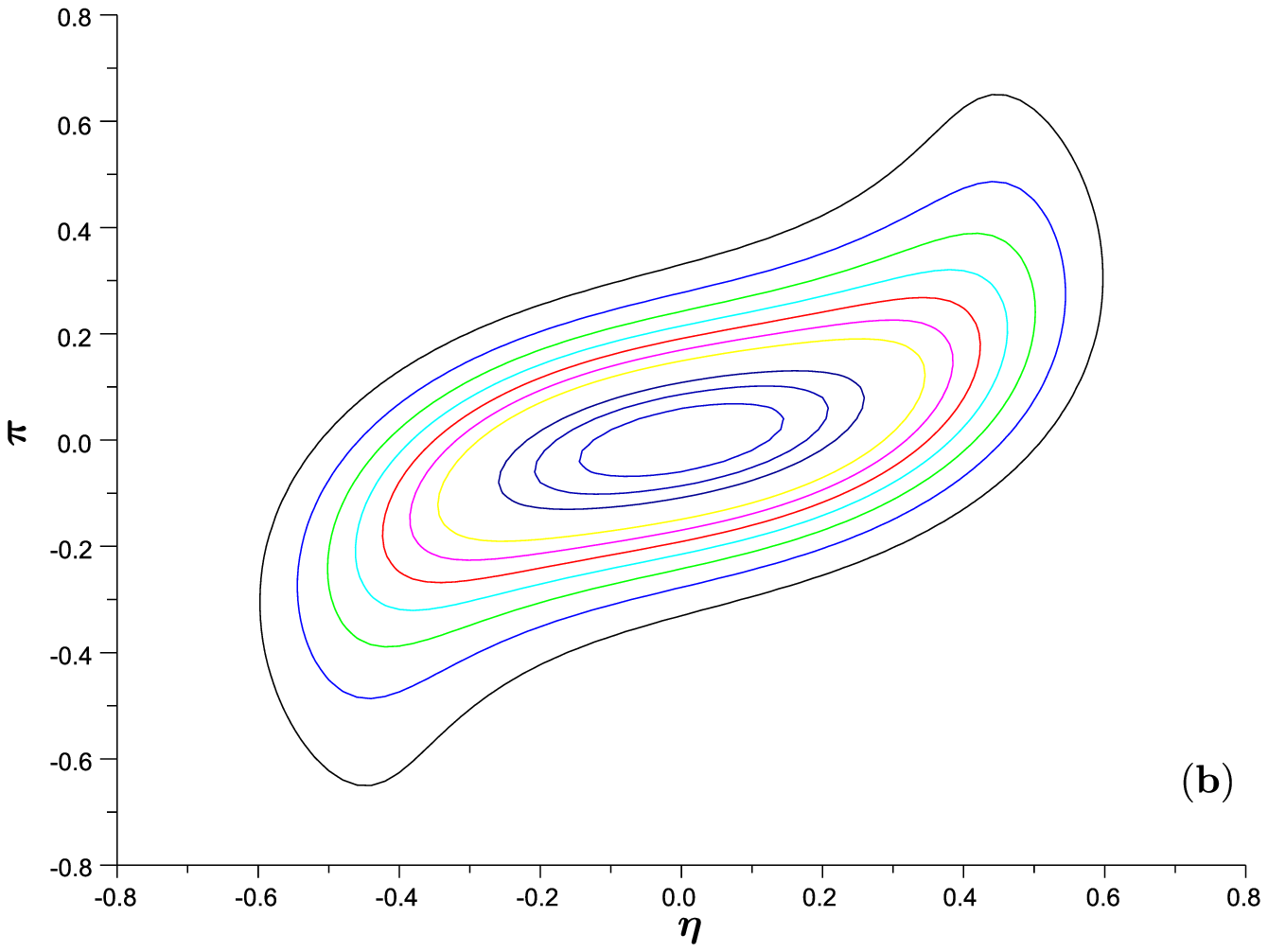} \\
\includegraphics[scale=0.42]{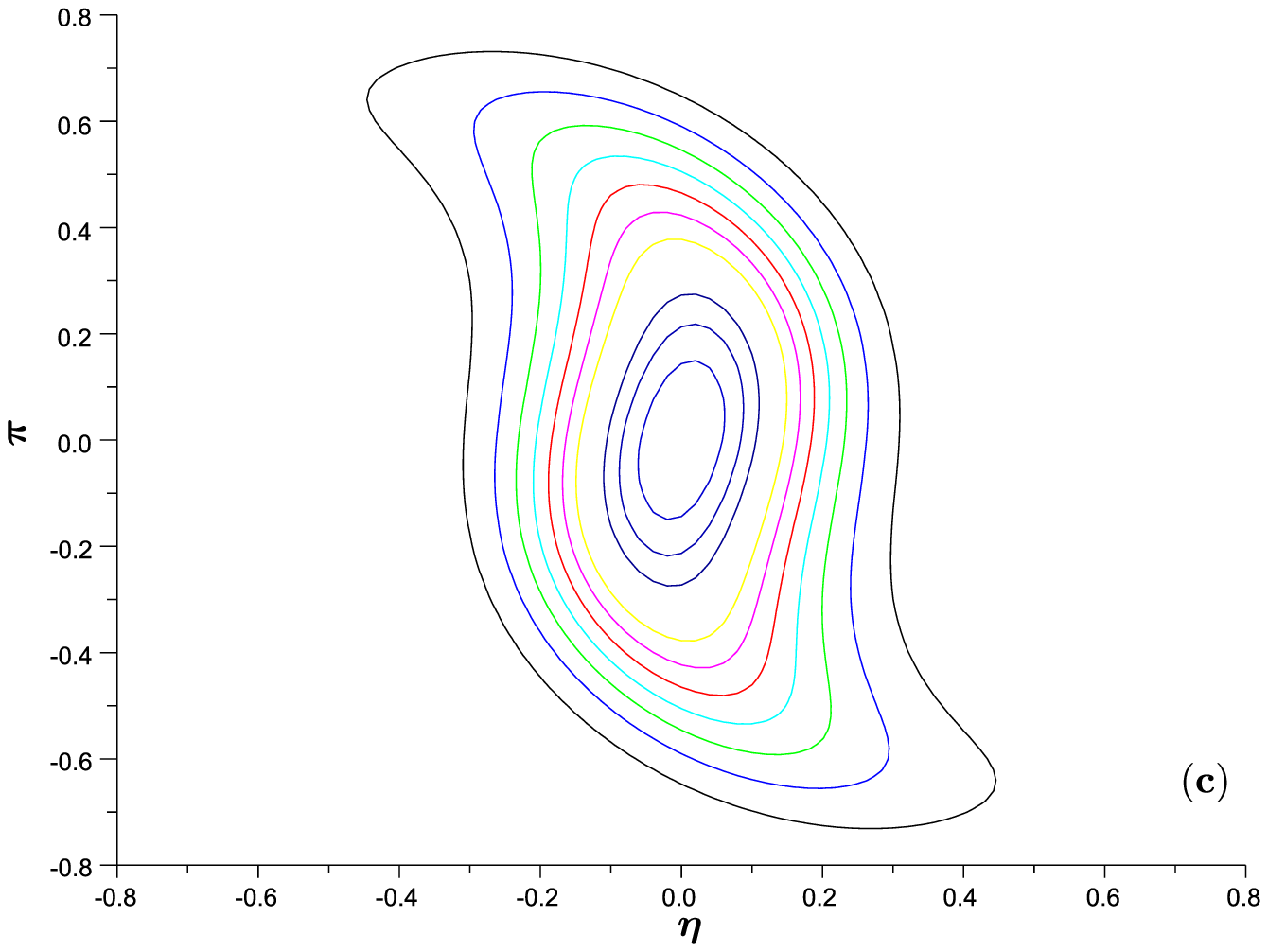}\includegraphics[scale=0.42]{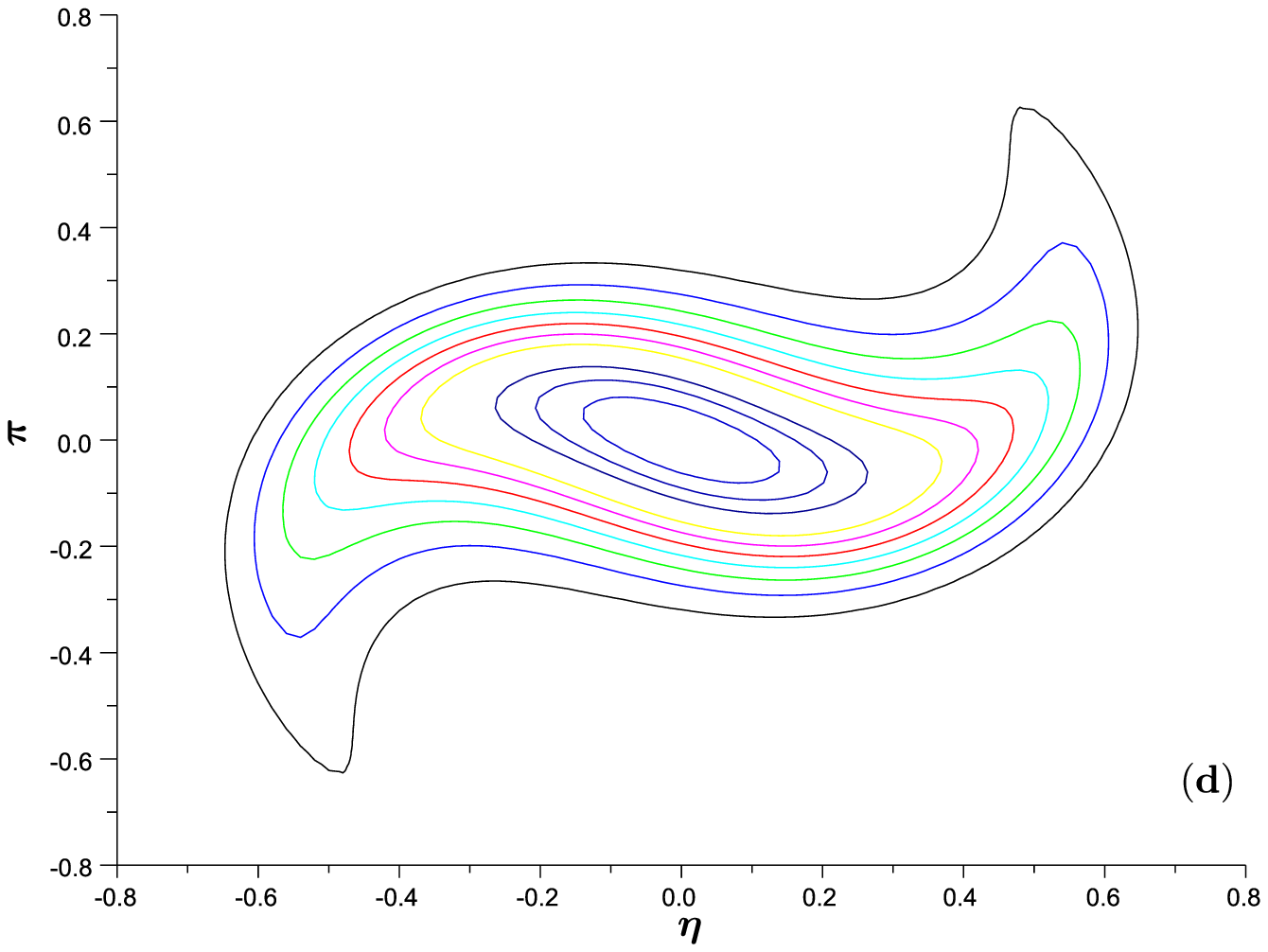}
\caption{Phase space contour plots after $n$ steps $\delta\lambda = 5 \times 10^{- 3}$: 
(a) $n = 0$; (b) $n = 1000$; (c) $n = 1400$; (d) $n = 2400$.}
\label{phspev}
\end{figure}

We consider the relativistic Hamiltonian
\begin{equation}
\label{Hamho}
H = c\,\sqrt{p^2 + (m_0\,c)^2}+\frac12\,k\,x^2
\end{equation}
which for future convenience will be written as
\begin{equation}
H = m_0\,c^2\,\left(\sqrt{1 + \Pi^2} + \frac12\,\eta^2\right)
\end{equation}
where
\begin{equation}
\eta = \frac{\Omega}{c}\,x, \qquad\qquad \Omega = \sqrt{\frac{k}{m_0}}.
\end{equation}
By setting $\lambda = \Omega\,t$, the associated Liouville operator reads
\begin{equation}
\hat{L} = -\frac{\Pi}{\sqrt{1 + \Pi^2}}\,\frac{\partial}{\partial \eta} + \eta\,\frac{\partial}{\partial \Pi}.
\end{equation}
and the use of the symmetric split method yields the iterated solution 
\begin{equation}
\rho_n = \rho (\eta_n, \Pi_n;  0),
\end{equation}
with ($\delta \lambda = \Omega\,\delta t$)
\begin{equation}
\eta_n = \eta_{n - 1} - F^\prime \left(\Pi_{n - 1} + \eta_{n - 1}\,\frac{\delta\lambda}2\right)\,\delta\lambda,
\qquad\qquad F (\Pi) = \sqrt{1 + \Pi^2},
\end{equation}
and
\begin{equation}
\label{Pin}
\Pi_n = \Pi_{n - 1} + \frac{\eta_n + \eta_{n - 1}}2\,\delta\lambda.
\end{equation}
Fig. \ref{phspev} show the time evolution in phase space of an initially Gaussian distribution.
\begin{figure}[htb]
\includegraphics[scale=0.8]{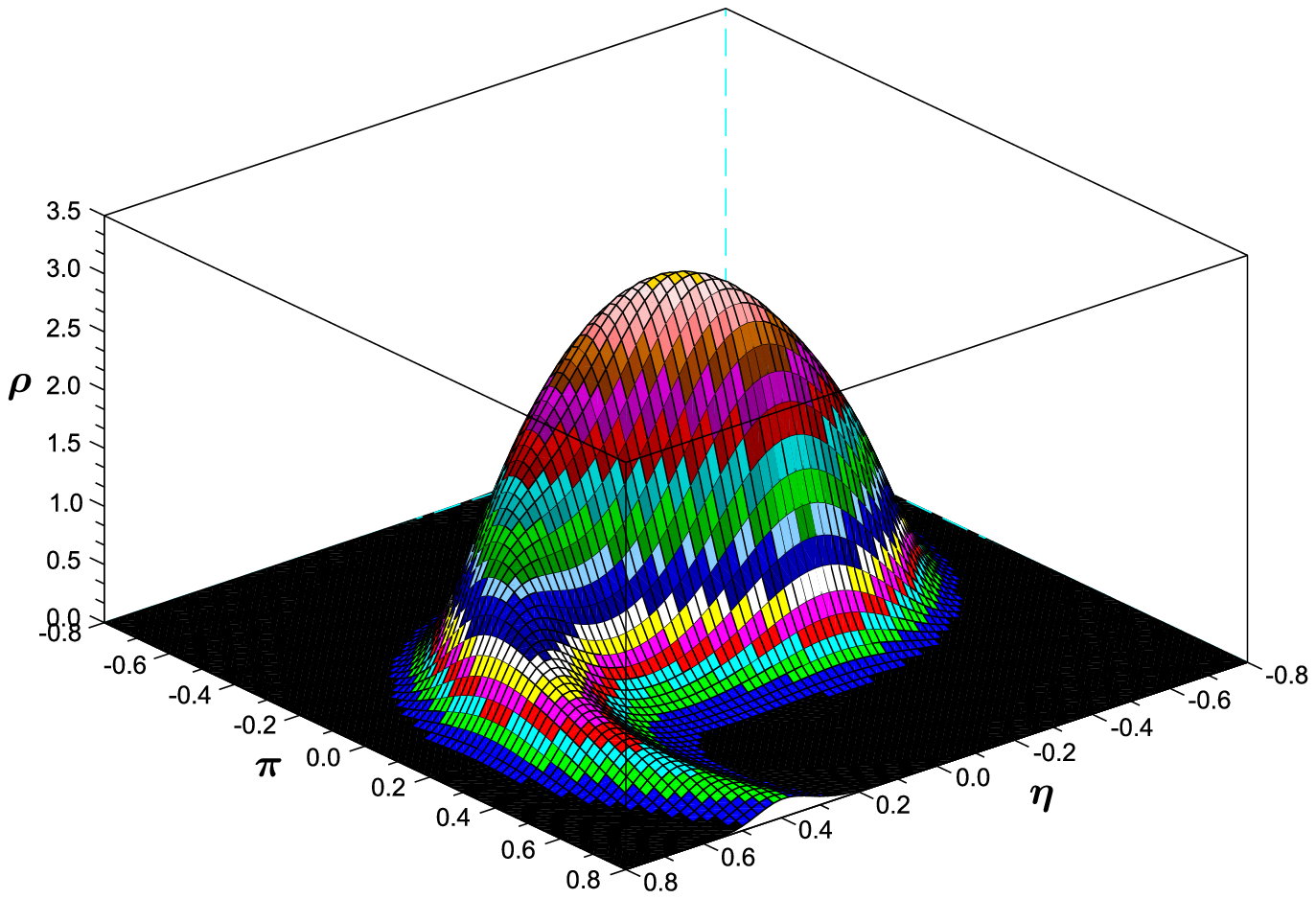}
\caption{Three-dimensional phase space density after 2400 steps.}
\label{3Dphsp}
\end{figure}

\begin{figure}[htb]
\includegraphics[scale=0.42]{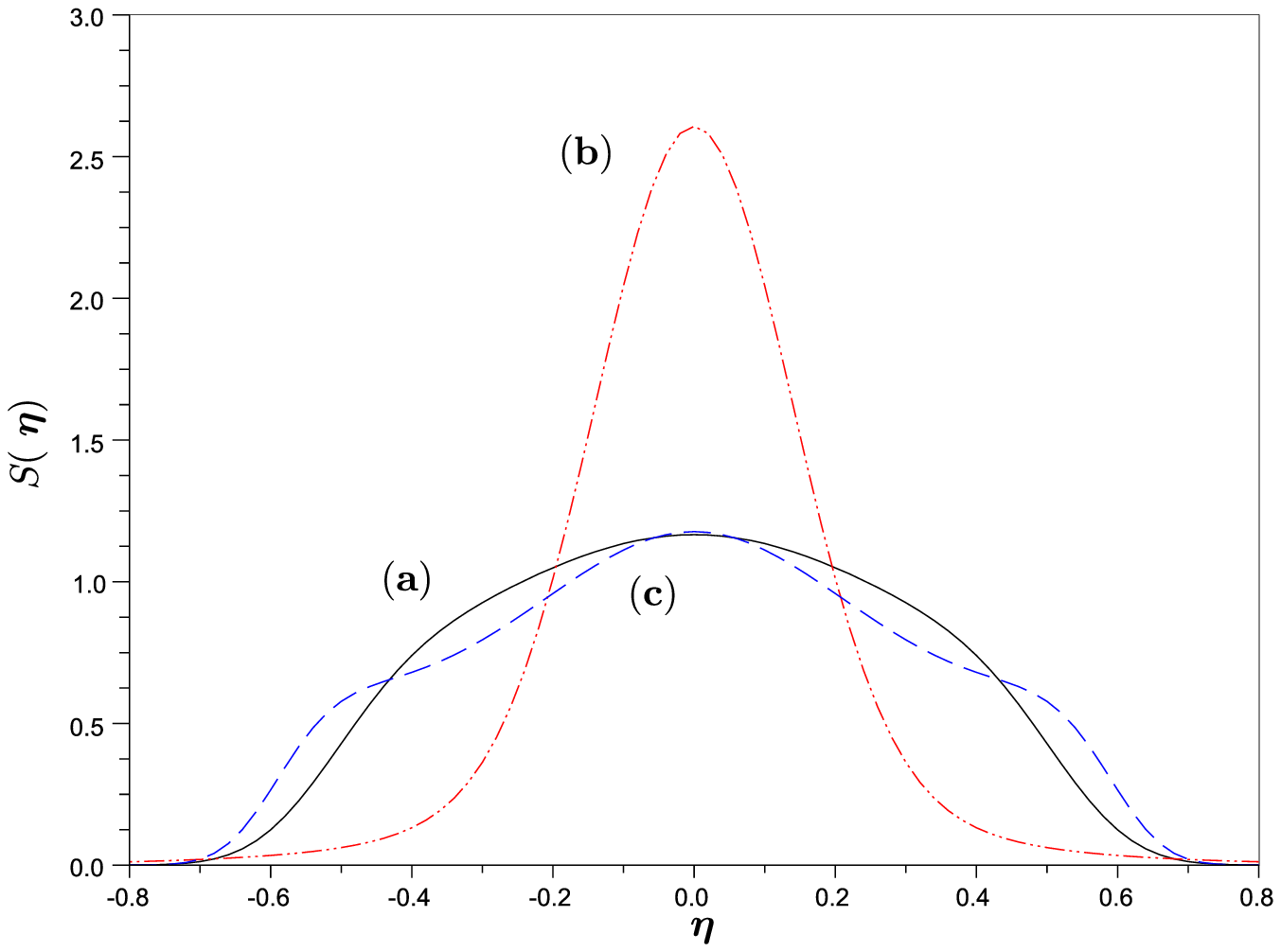}\includegraphics[scale=0.42]{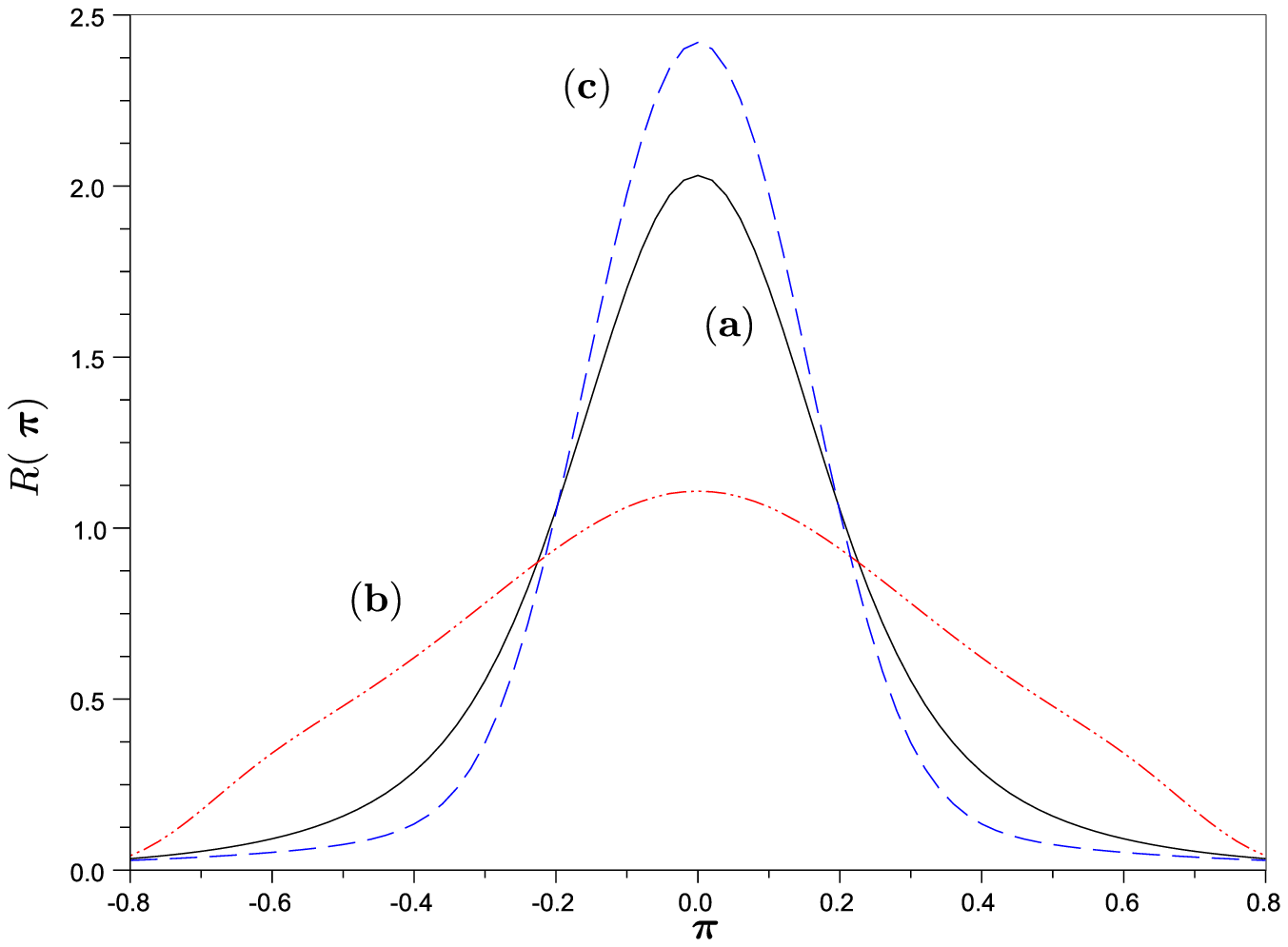}
\caption{The distributions defined in Eq. \eqref{SRdist} after $n$ steps $\delta\lambda = 5 \times 10^{- 3}$: 
(a) $n = 1000$; (b) $n = 1400$; (c) $n = 2400$.}
\label{distr}
\end{figure}

A global idea of distortion induced by relativistic effects in the distribution is offered by Fig. \ref{3Dphsp}, where we have reported the 3-dimensional plot after 2400 steps. The distortion is more 
significant on the tail of the distribution, where the momenta are larger and, thus, the relativistic effects are more effective. The evolution differs significantly from the corresponding non-relativistic 
case, where the distribution remains Gaussian and the evolution of the phase space contour plots are realized through continuous deformations of equal area ellipses. The distribution of spatial 
and momentum variables, defined by
\begin{equation}
\label{SRdist}
S_n (\eta) = \int_{-\infty}^\infty \mathrm{d}\Pi\,\rho_n, \qquad\qquad
R_n (\Pi) = \int_{-\infty}^\infty \mathrm{d}\eta\,\rho_n,
\end{equation}
are shown in Fig. \ref{distr}. 

\begin{figure}[htb]
\includegraphics[scale=0.42]{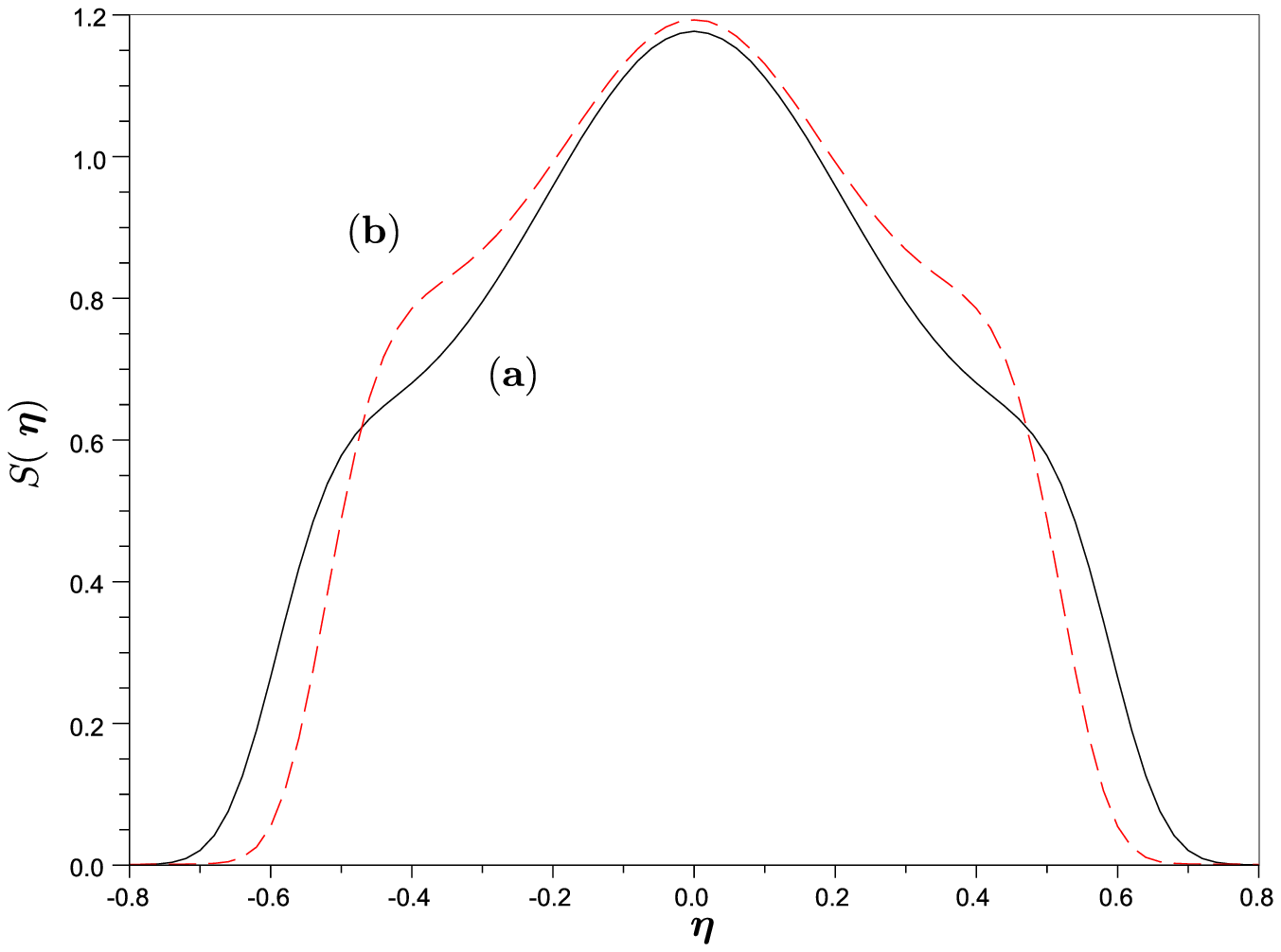}\includegraphics[scale=0.42]{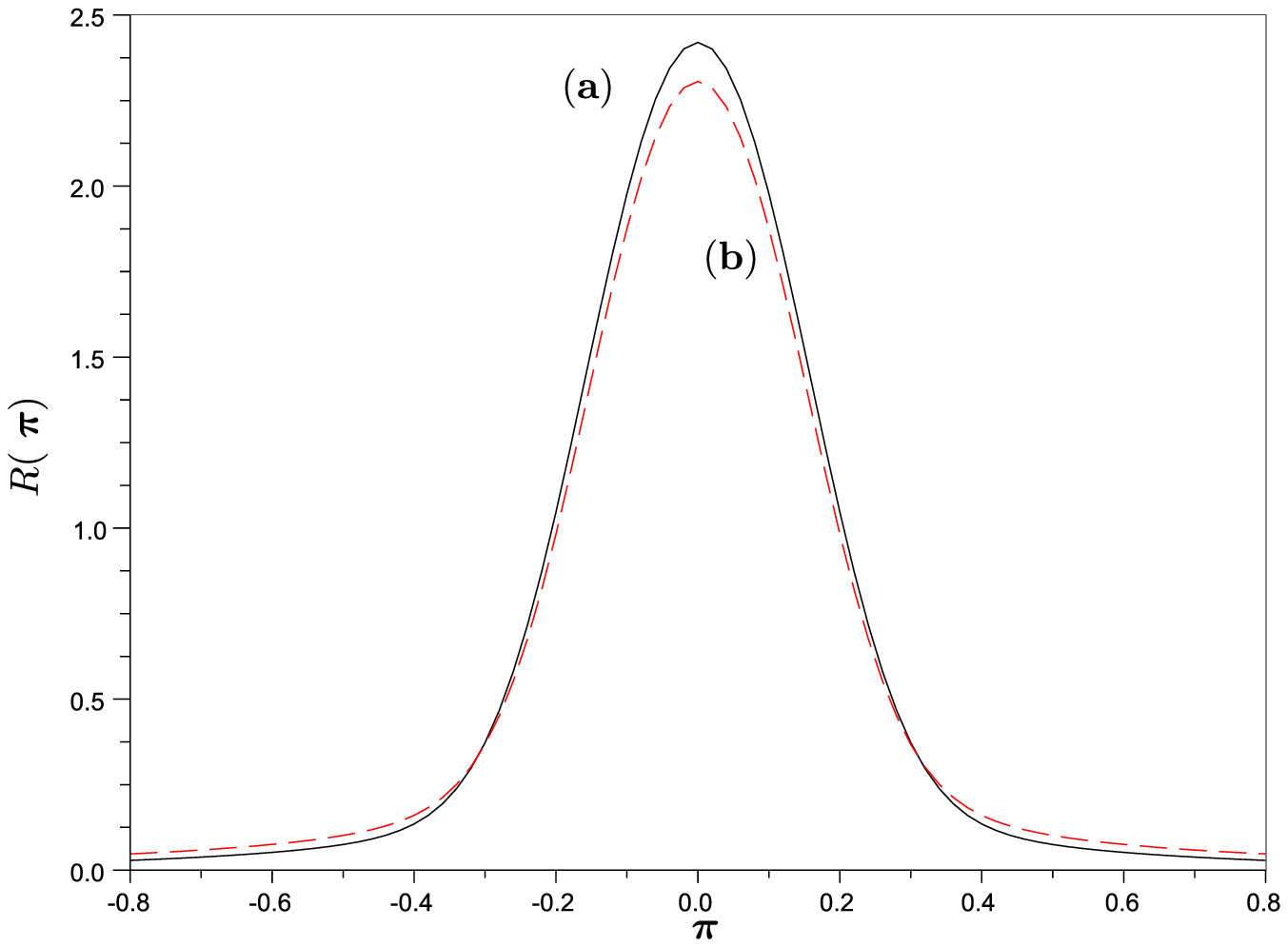}
\caption{The distributions of coordinate and momentum at the step $n = 2400$ for Hamiltonians \eqref{Hamho} (curve (a)) and \eqref{hoHam} (curve (b)).}
\label{spam2400}
\end{figure}

Let us now consider an analogous \eqref{Ham2} for the Hamiltonian of relativistic harmonic oscillator, namely
\begin{equation}
\label{hoHam}
H = \sqrt{(p\,c)^2 + \left(m_0\,c^2 + \frac12\,k\,x^2\right)^2}.
\end{equation}
We discuss this example in order to test the the efficiency of the method on a more complicated mathematical structure than the previous case, and to deal with a relativistic Hamiltonian 
containing higher order non-linearities in the spatial coordinate. Such a model Hamiltonian could, in principle, be exploited to study the relativistic motion of electrons in non-homogeneous magnetic 
fields \cite{Dattoli93}. The associated Liouville equation reads
\begin{equation}
\frac{\partial}{\partial \lambda} \rho = - \frac1{\sqrt{\Pi^2 + \left(1+ \displaystyle \frac12\,\eta^2\right)^2}}\,\left[\Pi\,\frac{\partial}{\partial \eta} \rho
- \eta\,\left(1+ \frac12\,\eta^2\right)\,\frac{\partial}{\partial \Pi} \rho\right]
\end{equation}
and the use of the symmetric split method yields the following iterated solution
\begin{equation}
\rho_n = \rho (\eta_n,\Pi_n; 0)
\end{equation}
with
\begin{equation}
\eta_n = \eta_{n - 1}  - \frac{\Pi_{n - 1} + \eta_{n - 1}\,\displaystyle \frac{\delta\lambda}2}{\sqrt{\left(\Pi_{n - 1} + \eta_{n - 1}\,\displaystyle \frac{\delta\lambda}2\right)^2 + 
\left(1 + \displaystyle \frac12\,\eta_{n - 1}^2\right)^2}}\,\delta\lambda
\end{equation}
and $\Pi_n$ given by Eq. \eqref{Pin}. 
In Fig. \ref{spam2400} the coordinate and momentum distributions for the Hamiltonians \eqref{Hamho} and \eqref{hoHam} are compared.

\section{Concluding remarks}
\begin{figure}[htb]
\includegraphics[scale=0.8]{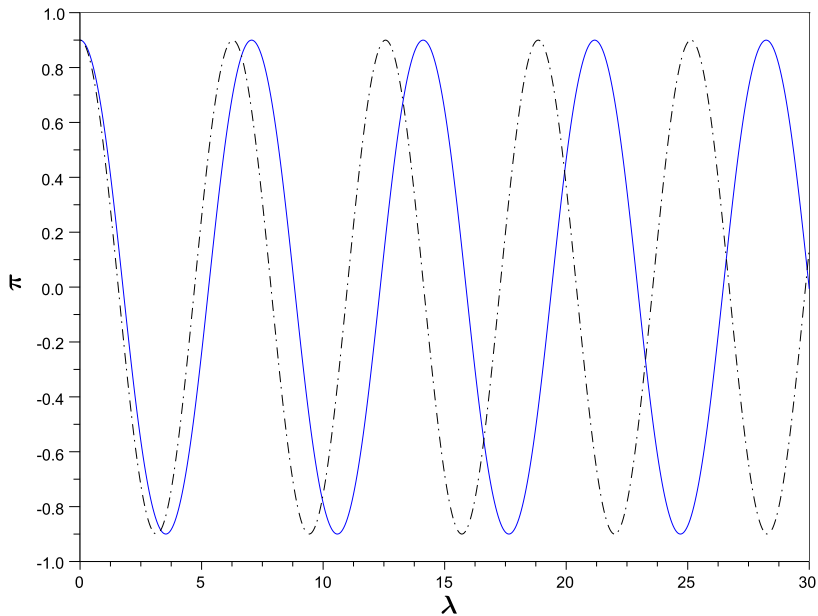}
\caption{Solution of Eq. \eqref{nleq} ($\Pi_0 = 0.9, \Pi_0^\prime = 0$) obtained with standard numerical methods and with the Lie series. 
The results coincide (blue continuous line). The dot-dashed line refers to the ordinary harmonic oscillator.}
\label{cmpLie}
\end{figure}

Even though we have insisted on the term relativistic ``harmonic oscillator", the relevant phenomenology is by no means harmonic. The interplay between energy and mass in relativistic dynamics 
induces non-linear terms that are responsible for anharmonic contributions in the equations of motion and can be, in principle, analytically accounted for by elliptic functions. This is not the case 
for the equation obtained from equation of motion associated to Hamiltonian \eqref{Hamho}, namely
\begin{equation}
\label{nleq}
\frac{\partial^2}{\partial \lambda^2} \Pi = -\frac{\Pi}{\sqrt{1 + \Pi^2}}.
\end{equation}
If we limit ourselves to the lowest-order corrections, we can expand this equation to get the following Duffing-type oscillator equation 
\begin{equation}
\Pi^{\prime\prime} = - \Pi + \frac12\,\Pi^3
\end{equation}
that is accurate up to $O(\beta^6)$, and whose solution is written, in terms of Jacobi functions, as \cite{Davis}
\begin{equation}
\label{Duff}
\Pi = \sqrt{2\,(1 - \sigma^2)}\,\mathrm{cn} \left(\sigma\,\lambda; \frac{\sigma^2 - 1}{2\,\sigma^2}\right)
\end{equation}
for the initial conditions 
\begin{equation}
\lambda = 0,\qquad \Pi_0 = \Pi (0) =  \sqrt{2\,(1 - \sigma^2)}, \qquad \Pi_0^\prime = \frac{\mathrm{d}}{\mathrm{d} \lambda} \Pi \mid_{\lambda = 0} = 0.
\end{equation}
The oscillation period of the function \eqref{Duff} is given by
\begin{equation}
T = \frac4{\Omega\,\sqrt{1 - \displaystyle \frac{\Pi_0^2}2}}\,K\left(\frac{\Pi_0^2}{2\,\Pi_0^2 - 4}\right) \cong \frac{2\,\pi}{\Omega}\,
\left[1 + \frac14\,\Pi_0^2\,\left(1 + \displaystyle \frac1{2\Pi_0^2 - 4}\right)\right]
\end{equation}
where $K$ is the complete elliptic integral of the first kind. On the r. h. s. we have reported the lowest order corrections, from which we may argue that even small kinetic energies (i.e.,  few keV for 
electrons) can induce corrections to the period of the order of 10\%. In Fig. \ref{cmpLie} we have reported a comparison between the solutions of Eq. \eqref{nleq} obtained by an integration with 
conventional means and via the symmetric split method. The results coincide and give complete confidence on the reliability of the previous analysis. On the other side the evolution operators 
associated with the Liouville equations or with the Hamiltonian counterparts are linked by  the relation $U_H (t) = U_L (-t)$ (see Ref. \cite{Dattoli95} for further comments). 

In this paper we have addressed the problem of phase space evolution for a relativistic system. We have used the Liouville equations, without taking any care about the transformation properties of 
the associated distribution. For this particular problem we have two different type of transformations: 
\begin{enumerate}
  \item [i)] those associated with the iterative scheme deriving from the symmetric split decomposition of the evolution operator,
  \item [ii)] the Lorentz transformation properties of the Liouville distribution. 
 \end{enumerate}
As for the first point, it is worth noting that they can be considered canonical transformations, for which, at each time step, we redefine the canonical coordinates 
$(\eta_n, \Pi_n) \to (\eta_{n + 1}, \Pi_{n + 1})$. It can be checked by direct computation it is verified the condition
\begin{equation}
J = \left|
\begin{array}{cc}
\dfrac{\partial \eta_n}{\partial \eta_{n - 1}} & \dfrac{\partial \eta_n}{\partial \Pi_{n - 1}}\\[10pt]
\dfrac{\partial \Pi_n}{\partial \eta_{n - 1}}  & \dfrac{\partial \Pi_n}{\partial \Pi_{n - 1}}
\end{array} 
\right| = 1
\end{equation}
that ensures the preservation of the norm of the distribution. 

\begin{figure}[htb]
\includegraphics[scale=0.8]{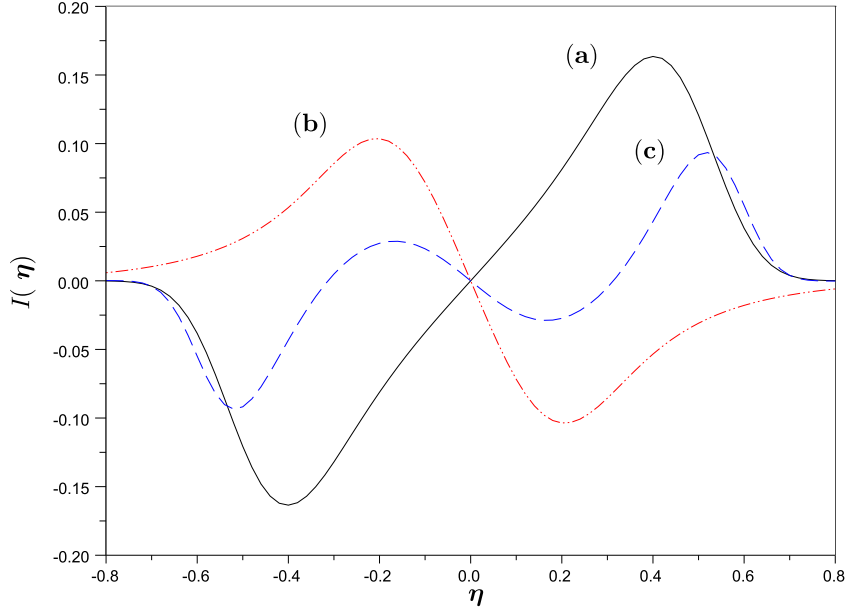}
\caption{Liouville density current for the relativistic harmonic oscillator $n$ steps $\delta\lambda = 5 \times 10^{- 3}$: a) $n = 1000$; 
b) $n = 1400$; c) $n = 2400$.}
\label{Lcur}
\end{figure}

Although the Liouville distribution is a Lorentz invariant (see Refs. \cite{Kampen,Bradt}), its spatial and momentum distributions do not possess this property. In particular, we can introduce the Liouville 
density current, defined as ($\eta^\prime = \mathrm{d} \eta/{\mathrm{d} \lambda}$)
\begin{equation}
S (\eta, \lambda) = \int_{-\infty}^\infty \mathrm{d}\Pi\,\rho (\eta, \Pi; \lambda), \qquad
I (\eta, \lambda) = \int_{-\infty}^\infty \mathrm{d}\Pi\,\eta^\prime\,\rho (\eta, \Pi; \lambda)
\end{equation}
transforming as a four-vector in Minkowski spacetime and satisfying the continuity equation
\begin{equation}
\label{cont}
\partial_\eta I (\eta,\lambda) + \partial_\lambda S (\eta,\lambda) = 0.
\end{equation}
We limited ourselves to a one dimensional oscillator, but the previous conclusions hold also in the three dimensional case. We have checked numerically the correctness of the continuity equation 
\eqref{cont} and in Fig. \ref{Lcur} we report the spatial and temporal parts of the Liouville density current for the harmonic oscillator.

In this paper we did not consider the problem from the quantum point of view, by taking into account the evolution of the Wigner distribution function, which can be considered the quantum analogous of the 
Liouville phase space density function. This problem has been considered in the literature (see Ref. \cite{Varro} and references therein), with particular emphasis on gauge invariant forms. The analysis we 
have developed here (in particular, the use of the split technique method) can also be extended to this case but completely new scenarios are open, mainly with reference to the equation ruling the evolution 
of the Wigner phase space distribution. The use of the Wigner-Moyal phase space operator representation \cite{Klauder} determines additional computational problems, which deserves more than a paragraph 
in a concluding section. It is our intention to consider the treatment of this problem in a forthcoming investigation.

A point that we can address, without taking too much space, is the solution of the relativistic Schr\"odinger equation for a spinless quantum relativistic particle under the influence of constant electric field.  From 
the mathematical point of view, the equation we should deal with is the following Salpeter equation \cite{Kowalski11}
\begin{equation}
i\,\partial_\tau \Psi (\xi,\tau) = \left[\sqrt{1 - \partial_\xi^2} - A\,\xi\right]\,\Psi (\xi, \tau), \qquad\qquad \Psi (\xi,0) = \psi (\xi).
\end{equation}
The same equation in the Fourier conjugated space reads
\begin{equation}
\label{Fou}
i\,\partial_\tau \tilde{\Psi} (\eta,\tau) = \left[\sqrt{1 + \eta^2} + i\,A\,\partial_\eta\right]\,\tilde{\Psi} (\eta,\tau),
\end{equation}
where
\begin{equation}
\tilde{\Psi} (\eta,\tau) = \frac1{\sqrt{2\,\pi}}\,\int_{-\infty}^\infty \mathrm{d}\xi\,e^{-i \eta \xi}\,\Psi (\xi, \tau).
\end{equation}
The solution of Eq. \eqref{Fou} writes
\begin{equation}
\tilde{\Psi} (\eta, \tau) = \exp\left\{-i\,\int_0^\tau \mathrm{d}\chi\,\sqrt{1 + (\eta - A\,\chi)^2}\right\}\,\tilde{\psi} (\eta - A\,\tau)
\end{equation}
and therefore its spatial counterpart can be written in terms of the following integral transform
\begin{equation}
\Psi (\xi, \tau) = \frac1{\sqrt{2\,\pi}}\,\int_{-\infty}^\infty \mathrm{d}\eta\,\exp\left\{i\,\left(\eta\,\xi -
 \int_0^\tau \mathrm{d}\chi\,\sqrt{1 + (\eta - A\,\chi)^2}\right)\right\}\,\tilde{\psi} (\eta - A\,\tau).
\end{equation}

In the case of a Salpeter equation with a quadratic potential, i.e
\begin{equation}
i\,\partial_\tau \Psi (\xi,\tau) = \left[\sqrt{1 - \partial_\xi^2} + B\,\xi^2\right]\,\Psi (\xi, \tau),
\end{equation}
by applying the symmetric split technique, and working in the conjugated space, we can get the solution in terms of the following recursion 
\begin{equation} 
\tilde{\Psi}_{n + 1} (\eta,\tau) = \frac{\exp\left\{-i\,\frac{\tau}2\,\sqrt{1 + \eta^2}\right\}}{2\,\sqrt{i\,\pi\,B\,\tau}}\,\int_{- \infty}^\infty \mathrm{d}\sigma\,
\exp\left\{- \frac{(\eta - \sigma)^2}{4\,i\,B\,\tau} + i\,\frac{\tau}2\,\sqrt{1 + \eta^2}\right\}\,\tilde{\Psi}_n (\sigma, \tau),
\end{equation}
($\tilde{\Psi}_0 (\eta,\tau) = \tilde{\Psi} (\eta, 0)$) where to describe the action of the transformed quadratic potential on the successive steps of the iteration we have used the following imaginary 
Weierstrass transform 
\begin{equation}
e^{i\,a\,\partial_\eta^2}\,f (\eta) = \frac1{2\,\sqrt{i\,\pi\,a}}\,\int_{- \infty}^\infty \mathrm{d}\sigma\,\exp\left\{- \frac{(\eta - \sigma)^2}{4\,i\,a}\right\}\,f (\sigma)
\end{equation}
This approach can be complemented with the generalized transform methods developed in Refs. \cite{Kowalski11} and \cite{ Kowalski07}.  As it will be shown elsewhere, this procedure can 
be efficiently used to study evolution of the wavefunction of a relativistic spinless particle.


\end{document}